\documentclass[english,aps,manuscript]{revtex4}
\usepackage[T1]{fontenc}
\usepackage[latin1]{inputenc}
\makeatletter

\usepackage{babel}
\makeatother
\begin{document}

\title{Neutrino mixings and magnetic moments due to Planck scale effects }

\author{Bipin Singh Koranga }

\address{Department of Physics, Indian Institute of Technology Bombay, Mumbai
400076, India.}

\begin{abstract}
In this paper, we consider the effect of Planck scale operators on
neutrino magnetic moments. We assume that the main part of neutrino
masses and mixings arise through GUT scale operators. We further assume
that additional discrete symmetries make the neutrino mixing bi-maximal.
Quantum gravitational (Planck scale) effects lead to an effective
$SU(2)_{L}\times U(1)$ invariant dimension-5 Lagrangian involving
neutrino and Higgs fields, which gives rise to additional terms in
neutrino mass matrix. These additional terms can be considered to
be perturbation of the GUT scale bi-maximal neutrino mass matrix.
We assume that the gravitational interaction is flavour blind and
we study the neutrino mixings and magnetic moments due to the physics
above the GUT scale.
\end{abstract}
\maketitle

\section{Introduction}

Neutrino magnetic moments is proportional to the neutrino mass as
required by the symmetry principles. At present the solar, atmospheric,
reactor and accelerator experiments indicates the existence of non
zero neutrino masses. Its indicates that neutrino has a magnetic moment.
A minimal extension of solar model yields a neutrino magnetic moment
{[}1{]}

\begin{equation}
\mu_{\nu}=\frac{3eG_{F}m_{\nu}}{8\pi^{2}\sqrt{2}}=\frac{3G_{F}m_{e}m_{\nu}}{4\pi^{2}\sqrt{2}}\mu_{B},\end{equation}

where $\mu_{B}=e/2m$ is the Bohr magnetron, $m_{e}$ is the electron
mass and $m_{\nu}$ is the neutrino mass.

The fundamental magnetic moment are associated with the mass eigenstates
in the mass eigenstates basis. Dirac neutrino can have diagonal or
off diagonal moment, while Majarona neutrino can have transition magnetic
moments {[}2, 3, 4{]}. The experimental value of neutrino magnetic
moment can be determined by only in the recoil electron spectrum from
neutrino electron spectrum {[}5, 6{]}. In this paper, we study, how
Planck scale effects the neutrino magnetic moments. Magnetic moment
of neutrinos, in principle, depend on the distance from its source
{[}4{]}

\begin{equation}
\mu_{e}^{2}=\sum_{i}\,\,|\sum_{j}U_{ej}\mu_{ij}exp(-iE_{j}L)|^{2},\end{equation}

where $\mu_{ij}$ is the fundamental constant in term of unit $\mu_{B}$
that characterize the coupling of the neutrino mass eigenstate to
the electromagnetic field. The expression for $\mu_{e}^{2}$ in the
case of Dirac neutrino, with only diagonal magnetic moment ($\mu_{ij}=\mu_{i}\delta_{ij})$;
this is used by the Particle Data Group {[}4{]}

\begin{equation}
\mu_{e}^{2}=\sum_{j}|U_{ej}|^{2}|\mu_{j}|^{2},\end{equation}

In this expression, there is no dependence of L and neutrino energy
E. In this one can say the neutrino magnetic moments depend on neutrino
mixings. In the case of Majarona neutrino, and we assume three mass
eigenstates. Then

\begin{equation}
\mu_{e}^{'^{2}}=(|\mu_{12}|^{2}+|\mu_{13}|^{2})(|U_{e2}|^{2}+|U_{e3}|^{2}).\end{equation}

For the Dirac case this implies that at least nondoagonal magnetic
moment is as large as the diagonal ones. In the case of Majorana,
it implies that two different nondiagonal magnetic moment are of a
similar magnitude {[}16{]}. Correction to neutrino mixing and neutrino
magnetic moments are given in Section 2. In section 3 give the results
on neutrino mixing and magnetic moments.

\section{Corrections to mixing angles and neutrino magnetic moments}

The neutrino mass matrix is assumed to be generated by the see saw
mechanism {[}7, 8, 9{]}. Here we will assume that the dominant part
of neutrino mass matrix arises due to GUT scale operators and they
lead to bi-maximal mixing. The effective gravitational interaction
of neutrinos with Higgs field can be expressed as $SU(2)_{L}\times U(1)$
invariant dimension-5 operator {[}10{]},

\begin{equation}
L_{grav}=\frac{\lambda_{\alpha\beta}}{M_{pl}}(\psi_{A\alpha}\epsilon_{AC}\psi_{C})C_{ab}^{-1}(\psi_{Bb\beta}\epsilon_{BD}\psi_{D})+h.c.\end{equation}

Here and every where below we use Greek indices $\alpha,\beta$..for
the flavour states and Latin indices i, j, k for the mass states.
In the above equation $\psi_{\alpha}=(\nu_{\alpha},l_{\alpha})$ is
the lepton doublet~, $\phi=(\phi^{+},\phi^{0})$ is the Higgs doublet
and $M_{pl}=1.2\times10^{19}GeV$ is the Planck mass. $\lambda\,\,$is
a $3\times3$ matrix in flavour space with each element O(1). In eq(4),
all indices are explicitly shown. The Lorentz indices a, b = 1, 2,
3, 4 are contracted with the charge conjugation matrix C and the SU(2)L
isospin indices A, B, C, D = 1, 2 are contracted with $\epsilon,$
the Levi-Civita symbol in two dimensions. After spontaneous electroweak
symmetry breaking the Lagrangian in eq(4) generates additional terms
of neutrino mass matrix

\begin{equation}
L_{mass}=\frac{v^{2}}{M_{pl}}\lambda_{\alpha\beta}\nu_{\alpha}C^{-1}\nu_{\beta},\end{equation}

where $v$=174 GeV is the VEV of electroweak symmetry breaking. 

We assume that the gravitational interaction is {}``flavour blind''
, that is $\lambda_{\alpha\beta}\,\;$ is independent of $\alpha,\,\,\beta$~indices.
Thus the Planck scale contribution to the neutrino mass matrix is

\begin{equation}
\mu\,\,\,\lambda=\mu\left(\begin{array}{ccc}
1 & 1 & 1\\
1 & 1 & 1\\
1 & 1 & 1\end{array}\right),\end{equation}

where the scale $\mu$~is

\begin{equation}
\mu=\frac{v^{2}}{M_{pl}}=2.5\times10^{-6}eV.\end{equation}

In our calculation, we take eq(6) as a perturbation to the main part
of the neutrino mass matrix, that is generated by GUT dynamics. We
compute the changes in neutrino mass eigenvalues and mixing angles
induced by this perturbation. We assume that GUT scale operators give
rise to the light neutrino mass matrix, which in mass eigenbasis,
takes the form $M=diag($$M_{1},$$M_{2}$, $M_{3})$, where $M_{i}$
are real and non negative. We take these to be the unperturbed ($0^{th}-order)$~masses.
Let $U$ be the neutrino mixing matrix at $0^{th}-order$. Then the
corresponding $0^{th}-order$ mass matrix $\mathbf{M}$ in flavour
space is given by

\begin{equation}
\mathbf{M}=U^{*}MU^{\dagger}.\end{equation}

The $0^{th}-order$ MNS matrix $U$ is given in this form 

\begin{equation}
U=\left(\begin{array}{ccc}
U_{e1} & U_{e2} & U_{e3}\\
U_{\mu1} & U_{\mu2} & U_{\mu3}\\
U_{\tau1} & U_{\tau2} & U_{\tau3}\end{array}\right),\end{equation}

where the nine elements are functions of three mixing angles, one
Dirac phase and two Majorana phases. In terms of the above elements,
the mixing angles are defined by 

\begin{equation}
|\frac{U_{e2}}{U{}_{e1}}|=tan\theta_{12},\,\,\,\,\frac{U_{\mu3}}{U_{\tau3}}|=tan\theta_{23},\,\,\,\,|U_{e3}|=sin\theta_{13}.\end{equation}

In terms of the above mixing angles, the mixing matrix is written
as 

\begin{equation}
U=diag(e^{if1},\,\, e^{if2},\,\, e^{if3})R(\theta_{23})\Delta R(\theta_{13})\Delta^{*}R(\theta_{12})diag(e^{ia1},e^{ia2},1).\end{equation}

The matrix $\Delta=diag(e^{\frac{i\delta}{2}},1,e^{\frac{-i\delta}{2}})$
contains the Dirac phase $\delta.$ This leads to CP violation in
neutrino oscillations. $a1$ and $a2$~are the so called Majorana
phases, which affect the neutrinoless double beta decay. $f1,\,\, f2$
and $f3$~are usually absorbed as a part of the definition of the
charge lepton field. It is possible to rotate these phases away, if
the mass matrix eq(5) is the complete mass matrix. However, since
we are going to add another contribution to this mass matrix, these
phases of the zeroth order mass matrix can have an impact on the complete
mass matrix and thus must be retained. By the same token, the Majorana
phases which are usually redundant for oscillations have a dynamical
role to play now. Planck scale effects will add other contributions
to the mass matrix. Including the Planck scale mass terms, the mass
matrix in flavour space is modified as

\begin{equation}
\mathbf{M}\rightarrow\mathbf{M^{'}=M}+\mu\lambda,\end{equation}

with $\lambda$ being a matrix whose elements are all 1 as discussed
in eq(3). Since $\mu$ is small, we treat the second term (the Planck
scale mass terms) in the above equation as a perturbation to the first
term (the GUT scale mass terms). The impact of the perturbation on
the neutrino masses and mixing angles can be seen by forming the hermitian
matrix

\begin{equation}
\mathbf{M^{'^{\dagger}}M^{'}=}(\mathbf{M}+\mu\lambda)^{\dagger}(\mathbf{M}+\mu\lambda),\end{equation}

which is the matrix relevant for oscillation physics. To the first
order in the small parameter $\mu$, the above matrix is

\begin{equation}
\mathbf{M}^{\dagger}\mathbf{M}+\mu\lambda^{\dagger}\mathbf{M}+\mathbf{M}^{\dagger}\mu\lambda.\end{equation}

This hermitian matrix is diagonalized by a new unitary matrix $U^{'}.$
The corresponding diagonal matrix $M^{'^{2}},$ correct to first order
in $\mu$, is related to the above matrix by $U^{'}M^{'^{2}}U^{'^{\dagger}}.$
Rewriting $\mathbf{M}$ in the above expression in terms of the diagonal
matrix $M$ we get 

\begin{equation}
U^{'}M^{'^{2}}U^{'^{\dagger}}=U(M^{2}+m^{\dagger}M+Mm)U^{\dagger}\end{equation}

where \begin{equation}
m=\mu U^{t}\lambda U.\end{equation}

Here $M$ and $M^{'}$ are the diagonal matrices with neutrino masses
correct to $0^{th}\,\,$and $1^{th}\,\,$order in $\mu$. It is clear
from eq(15) that the new mixing matrix can be written as:

\[
U^{'}=U(1+i\delta\Theta),\]

\[
=\left(\begin{array}{ccc}
U_{e1} & U_{e2} & U_{e3}\\
U_{\mu1} & U_{\mu2} & U_{\mu3}\\
U_{\tau1} & U_{\tau2} & U_{\tau3}\end{array}\right)+\]

\begin{equation}
\left(\begin{array}{ccc}
U_{e2}\delta\Theta_{12}^{*}+U_{e3}\delta\Theta_{23}^{*}, & U_{e1}\delta\Theta_{12}+U_{e3}\delta\Theta_{23}^{*}, & U_{e1}\delta\Theta_{13}+U_{e3}\delta\Theta_{23}^{*})\\
U_{\mu2}\delta\Theta_{12}^{*}+U_{\mu3}\delta\Theta_{13}^{*}, & U_{\mu1}\delta\Theta_{12}+U_{\mu3}\delta\Theta_{23}^{*}, & U_{\mu1}\delta\Theta_{13}+U_{\mu3}\delta\Theta_{23}^{*}\\
U_{\tau2}\delta\Theta_{12}^{*}+U_{\tau3}\delta\Theta_{13}^{*}, & U_{\tau1}\delta\Theta_{12}+U_{\tau3}\theta\Theta_{23}^{*}, & U_{\tau1}\delta\Theta_{13}+U_{\tau3}\delta\Theta_{23}^{*}\end{array}\right),\end{equation}

where $\delta\theta$~is a hermitian matrix that is first order in
$\mu.$ From eq(15) we obtain

\begin{equation}
M^{2}+m^{\dagger}M+Mm=M^{'^{'2}}+[i\delta\Theta,M^{'^{2}}].\end{equation}

Therefore to first order in $\mu$,~the mass squared difference $\Delta M_{ij}^{2}=M_{i}^{2}-M_{j}^{2}$~get
modified {[}11, 13{]} as:

\begin{equation}
\Delta M_{ij}^{'^{2}}=\Delta M_{ij}^{2}+2(M_{i}Re[m_{ii}]-M_{j}Re[m_{jj}]).\end{equation}

The change in the elements of the mixing matrix, which we parametrized
by $\delta\Theta$, is given by

\begin{equation}
\delta\Theta_{ij}=\frac{iRe(m_{ij})(M_{i}+M_{j})}{\Delta M_{ij}^{'^{2}}}-\frac{Im(m_{ij})(M_{i}-M_{j})}{\Delta M_{ij}^{'^{2}}}.\end{equation}

The above equation determines only the off diagonal elements of matrix
$\delta\Theta_{ij}$. The diagonal elements of $\delta\Theta$ can
be set to zero by phase invariance.

The new Majorana neutrino magnetic moments due to Planck scale is
given by

\begin{equation}
\mu_{x}^{'^{2}}=\sum_{j}\sum_{k}|U_{xj}^{'}|^{2}|\mu_{jk}|^{2},\end{equation}

where ($x=e,\,\,\mu\,\,\tau)$ is the flavour indices. In the case
of three flavour, the magnetic moment of Majorana electron neutrinos
is given by

\begin{equation}
\mu_{e}^{'^{2}}=(|\mu_{12}|^{2}+|\mu_{13}|^{2})(|U_{e2}^{'}|^{2}+|U_{e3}^{'}|^{2}).\end{equation}

and there is no dependence on the distance L or neutrino energy.

\section{Results and Discussions}

We assume the largest allowed value of 2 eV for degenerate neutrino
mass which comes from tritium beta decay {[}12{]}. We also assume
normal neutrino mass hierarchy. Thus we have $M_{1}=$2 eV, $M_{2}=\sqrt{M_{1}^{2}+\Delta_{21}}$
and $M_{3}=\sqrt{M_{1}^{2}+\Delta_{31}}$. As in the case of $0^{th}$order
mixing angles, we can compute $1^{st}$ order mixing angles in terms
of $1^{st}$ order mixing matrix elements {[}14{]}. We expect the
mixing angles coming from GUT scale operators to be determined by
some symmetries. For simplicity, here we assume a bi-maximal mixing
pattern, $\theta_{12}=\theta_{23}=\pi/4$ and $\theta_{13}=0$. We
compute the modified mixing angles for the degenerate neutrino mass
of 2 eV.~ We have taken $\Delta_{31}=0.0025eV^{2}$ and $\Delta_{21}=0.00008eV^{2}$
. For simplicity we have set the charge lepton phases $f1=f2=f3=0$.
We have checked that non-zero values for these phases do not change
our results. Since we have set $\theta_{13}=0$, the Dirac phase $\delta$
drops out of the $0^{th}$ order mixing matrix. We consider the Planck
scale effects on neutrino mixing and we get the given range of mixing
parameter of MNS matrix

\begin{equation}
U^{'}=R(\theta_{23}+\epsilon_{3})U_{phase}(\delta)R(\theta_{13}+\epsilon_{2})R(\theta_{12}+\epsilon_{1}),\end{equation}

In Planck scale, only $\theta_{12}$($\epsilon_{1}=\pm3^{o})$have
reasonable deviation and $\theta_{13},\,\,\theta_{23}$ deviation
is very small less than $0.3^{o}$ {[}14{]}. In the new mixing at
Planck scale we get the given moments of Majorana neutrinos

\begin{equation}
\mu_{e}^{'^{2}}=(|\mu_{12}|^{2}+|\mu_{13}|^{2})(|U_{e2}^{'}|^{2}+|U_{e3}^{'}|^{2}).\end{equation}

\begin{equation}
\mu_{\mu}^{'^{2}}=(|\mu_{21}|^{2}+|\mu_{23}|^{2})(|U_{\mu1}^{'}|^{2}+|U_{\mu3}^{'}|^{2}).\end{equation}

\begin{equation}
\mu_{\tau}^{'^{2}}=(|\mu_{31}|^{2}+|\mu_{32}|^{2})(|U_{e\tau2}^{'}|^{2}+|U_{e\tau3}^{'}|^{2}).\end{equation}

The best direct limit on the neutrino magnetic moment, $\mu_{e}\leq1.8\times10^{-10}\mu_{B}$
at 90\% CL {[}15{]}, coming from neutrino electron scattering with
anti-neutrino. However, the limit obtained using the SK data {[}4{]},
$\mu_{e}\leq1.5\times10^{-10}\mu_{B}.$Due to Planck scale effects,
mixing angle $\theta_{12}$ and $\theta_{13}$ will contributes the
magnetic moments of neutrinos.

\section{conclusions}

We assumed that the main part of neutrino masses and mixings arise
from GUT scale operators. We considered these to be $0^{th}$ order
quantities. We further assumed that GUT scale symmetries constrain
the neutrino mixing angles to be either bi-maximal or tri-bi-maximal.
The gravitational interaction of lepton fields with SM Higgs field
gives rise to an $SU(2)_{L}\times U(1)$ invariant dimension-5 effective
lagrangian, given originally by Weinberg {[}10{]}. On electroweak
symmetry breaking this operator leads to additional mass terms. We
consider these to be a perturbation of GUT scale mass terms. We compute
the first order corrections to neutrino mass eigenvalues and mixing
angles. In {[}11{]}, it was shown that the change in $\theta_{13}$,
due to this perturbation, is small. Here we show that the change in
$\theta_{23}$ also is small (less than $0.3^{o})$ but the change
in $\theta_{12}\,\,$can be substantial (about $\pm3^{o})$. The changes
in all three mixing angles are proportional to the neutrino mass eigenvalues.
To maximize the change we assumed degenerate neutrino masses $\simeq2.0$~eV.
For degenerate neutrino masses, the changes in $\theta_{13}$~and
$\theta_{23}\,\,$are inversely proportional to $\Delta_{31}$ and
$\Delta_{32}$ respectively, whereas the change in $\theta_{12}$
is inversely proportional to $\Delta_{21}$. Since $\Delta_{31}\cong\Delta_{32}\gg\Delta_{21},$
the change in $\theta_{12}$ is much larger than the changes in $\theta_{13}\,\,$and
$\theta_{23}$. In this paper, we write the neutrino magnetic moment
expression for three flavour neutrino mixing. For majarona neutrino
two non diagonal moment, these expression are Eq(4.0) for vacuum mixing.
For majorana neutrino with three flavour, the expression is $\mu_{e}^{'^{2}}=(|\mu_{12}|^{2}+|\mu_{13}|^{2})(|U_{e2}|^{2}+|U_{e3}|^{2}).$
In this paper, finally we wish make a important comment. Due to Planck
scale effects mixing angle $\theta_{12}$ and $\theta_{13}$ contribute
the magnetic moments of neutrinos.

\end{document}